\documentclass[prl,aps,amssymb,amsmath,twocolumn,showpacs]{revtex4}
\newcommand{\eq}[1]{Eq.~{\ref{#1}}}
\newcommand{\fig}[1]{Fig.~{\ref{#1}}}
\usepackage{graphicx,epsfig}
\usepackage{dcolumn}
\begin{document}
\title{Width distribution
 of  contact lines on a disordered substrate}
\author{S\'ebastien Moulinet$^1$\footnote{now at LPMCN, Universit\'e
Claude Bernard, 43 Boulevard du 11 novembre 1918, 69622 Villeurbanne
Cedex, France}, 
 Alberto Rosso$^2$,
Werner Krauth$^1$, Etienne Rolley$^1$ } \affiliation{
$^{1}$CNRS-Laboratoire de Physique Statistique de l'Ecole Normale
Sup\'{e}rieure, 24 rue Lhomond, 75231 Paris, France\\
$^2$Universit\'e de Gen\`eve, DPMC,
24 Quai Ernest Ansermet, CH-1211 Gen\`eve 4, Switzerland
}
\date{\today}
\begin{abstract}
We have studied the roughness of a contact line of a liquid meniscus on a
disordered substrate by measuring its width distribution. The comparison
between the measured width distribution and the width distribution
calculated in previous works, extended here to the case of open boundary
conditions, confirms that the Joanny-de~Gennes model is not sufficient
to describe the dynamics of  contact lines at the depinning threshold.
This conclusion is in agreement with recent measurements which determine
the roughness exponent by extrapolation to large system sizes.
\end{abstract}


\maketitle

The physics of elastic interfaces in random media is involved in a vast class
of problems, such as domain walls in ferromagnetic
\cite{lemerle_domainwall_creep} or
 ferroelectric \cite{tybell_ferro_creep} systems,
and propagation of cracks in solids \cite{gao_crack_elasticity}.
A notorious example of  an elastic interface
 is provided by the  contact line
of a liquid meniscus  on a disordered substrate 
\cite{joanny-degennes,pomeau_contact_line,
joanny-robbins_contactline,ertas_long}.

In the past years, much effort has been devoted to shed light on  both the
 equilibrium properties and  the dynamics of this system
\cite{hazareezing,golestanian,prevost}, which is characterized by long-range
interactions.
Very  recently,  experiments  with  water or Helium
moving on a substrate characterized by a well controlled  disorder
\cite{prevost,moulinet_contact_line},  have explored the depinning threshold.
In this regime \cite{nattermann_stepanow_depinning,narayan_fisher_depinning}
the contact line, driven by an external force, moves very slowly.
The study of the roughness of the interface,
and in particular its scaling behavior,
turns out to be  a fundamental tool to test our understanding  of
the physics of systems in which  the elasticity and the disorder
compete in determining the shape of the interface.

\begin{figure}[t]
\includegraphics[width=\columnwidth]{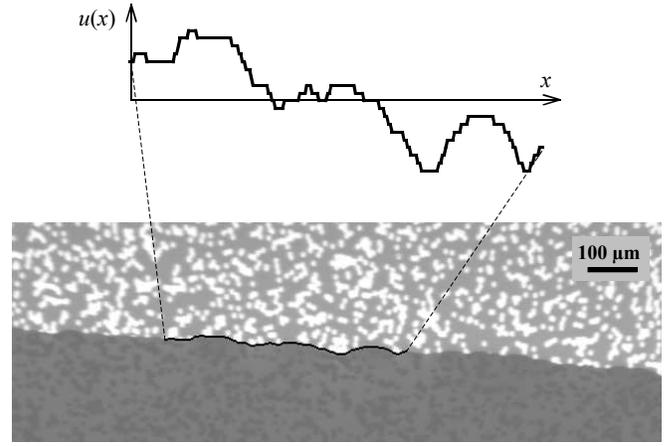}
\caption{Bottom: image of a water meniscus receding on a
disordered substrate. The wet region is darker. Chromium defects
appear as clear dots on the dry region. Top: sample of size $L=500
\, \mathrm{\mu m}$ extracted from  the digitalized contact line.
$u(x)$ is the deviation from the  mean height averaged over the
sample.} \label{f:cl}
\end{figure}

In \fig{f:cl} we display an experimental sample:
a glass plate with Chromium impurities 
 (clear dots), partially covered by a water meniscus
(dark region). The so-called contact line is  separating the wet  and
dry regions.
The contact line is defined by  an internal coordinate $x$ and
by the height $h$ along the motion direction.
We observe that a single-valued height function $h(x)$
is sufficient to characterize the shape of the contact line.
A sample of size $L=500 \, \mathrm{\mu m}$  extracted from  the image
is also shown in \fig{f:cl}; to analyze its  geometric properties
we are interested in the  deviations from the mean height
$u(x)=h(x)-\langle h \rangle$, with $\langle \ldots\rangle= 1 /L \int_0^L dx
\ldots$. The mean square width is defined as $w^2 = \langle u^2(x) \rangle$.
The roughness exponent $\zeta$ is introduced by considering an ensemble
of  lines of size $L$, where $L$ must be  larger than the
lengths which characterize the disorder.
Averaging over this ensemble,  we obtain $\overline{w^2} \propto L^{2\zeta}$
\cite{nattermann_stepanow_depinning}.

The theoretical evaluation of the exponent $\zeta$ has been
up to now based on the assumption
that  the motion of the line at the threshold is quasi-static
 \cite{narayan_fisher_depinning}; this assumption was shown to be valid for most viscous fluids \cite{moulinet_contact_line}.
This means that the equation of motion of the height  $h(x)$ can be derived
from an energy function, which incorporates the potential energy due to the
driving force $f$ and the disorder potential $\eta(x,h)$, as well as an
elastic energy.  According to this hypothesis,
the equation of motion for $h(x)$  at zero temperature is:
\begin{equation}
\frac{ \partial  }{\partial t } h(x)
=f + \eta(x,h) -k \int dx_{1}
\frac{h(x)-h(x_{1})}{(x-x_{1})^2} \,.
\label{motion}
\end{equation}
The last term in \eq{motion}
accounts for the  long-range elastic force
calculated  by Joanny and de~Gennes \cite{joanny-degennes}.
At the  equilibrium, independent  approaches
within \eq{motion} led  consistently to the value
$\zeta=1/3$ \cite{robbins_euro,hazareezing,ertas_long}.
At the depinning threshold, the determination of the roughness exponent
stimulated a large debate
\cite{ertas_long,narayan_fisher_depinning,tanguy,ramanathan}: finally,
extended  renormalization group calculations
up to the two-loop order proved that
$\zeta$ is larger than $1/3$ \cite{chauve_2loop}.
This finding  was confirmed  by a numerical study by means of an
exact algorithm, which is able to detect directly the blocked interface at
the depinning threshold: the precise resulting value
is $\zeta=0.388 \pm 0.002$ \cite{rosso_krauth_longrange}.

In spite of the large amount of theoretical work devoted to the
subject, experiments are very few (see \cite{johnsonD93} and
references therein). The main difficulty consists in  controlling
the disorder; in Ref.~\cite{moulinet_contact_line} this difficulty
has been overcome by using photolithographic techniques. Small
squares of Chromium (size: $10 \times 10\, \mathrm{\mu m}^2$) were
deposited randomly on a glass plate,  such that the  $22\%$ of the
surface is  covered. This procedure generates a disorder
correlated on a scale $\xi \sim 10\,\mathrm{\mu m}$,
 sufficiently large to prevent
thermal fluctuations from  playing any role.
The correlation length $\xi$ is
more than two orders of magnitude below the capillary length $L_c$
($\sim 2.5 \text{mm}$ in this system) where gravity begins to
limit the fluctuations. 
 When the experiment
is carried out, the glass plate is withdrawn very slowly from
the liquid bath at a  fixed velocity ranging between $0.2$ and
$20\,\mathrm{\mu m/s}$. The liquid is pure water or an aqueous solution of
glycerol with a  viscosity up to $20$ times that 
of water. One observes that  the shape of the
contact line is independent of the velocity. This is a clear
signature of the  depinning limit \cite{moulinet_contact_line}.
From these measurements it has been obtained, for $2\xi \le L \le
L_c/2$, $\zeta=0.51 \pm 0.03$, in disagreement with all
theoretical predictions \cite{rosso_krauth_longrange}.

The discrepancy between the theoretical and the measured roughness exponent
suggest that a richer model, more complicated than the one described by
 \eq{motion}, is needed to account for the critical behavior of
the contact line at the depinning.
However, as the range of accessible scales is less than two orders
 of magnitude,
the rigorous determination of the  exponent $\zeta$  is
a very delicate experimental task.
Thus, to reach  more convincing conclusions, it is desirable
to compare some other universal quantities characterizing
the depinning threshold. In particular, special care must be devoted
to evaluate finite size effects, which could affect drastically the
final results. To this purpose, we study here the complete
width distribution $P(w^2)$, and we compare our theoretical
predictions, based on the solution of \eq{motion},
to experimental data.

 The theoretical studies performed on systems without disorder such 
 as stochastic
models \cite{racz.random.94,racz.growth.94,plischke.curvature.94} or
magnetic systems \cite{bramwell.xy} have shown that it is possible to express
 $P(w^2)$ in a universal  scaling form:
\begin{equation}
 P({w^2}) = \frac{1}{\overline{w^2}}\phi (x = w^2/\overline{w^2}),
\label{e:scaling}
\end{equation}
where $x$ is the  renormalized width $w^2/\overline{w^2}$.
In \eq{e:scaling}  the size
dependence appears only through the average $\overline{w^2}$:  the
non-trivial scaling function $\phi(x)$ is  universal and  characterizes
a full class of systems.  Recently \cite{rosso_width}, the same scaling
properties have been proved for  elastic interfaces in random media
at the depinning threshold.  The width distribution of these systems is,
for all intents and purposes, given by a generalized Gaussian approximation of
independent modes, which decay with a characteristic propagator $G(q)
\sim q^{d+2\zeta}$.
Within this approximation, for interfaces of a fixed internal dimension
and for the required boundary conditions, $\phi(x)$ depends only
on the roughness exponent $\zeta$.

\begin{figure}[t]
\includegraphics[width=\columnwidth]{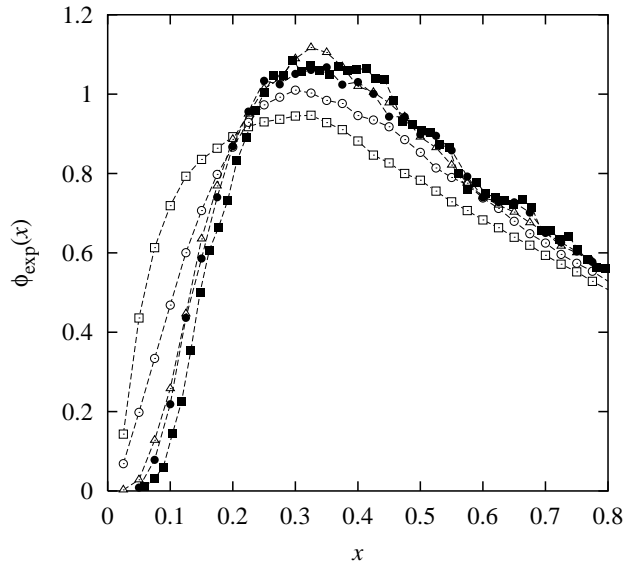}
\caption{Experimental scaling function $\phi_{\text{exp}}(x)$:
 study of finite size effects.
Open symbols: Aqueous solution of glycerol, $v=1 \,
 \mathrm{\mu m/s}$
(square: $L=124 \, \mathrm{\mu m}$, circle: $L=186 \, \mathrm{\mu m}$,
triangle: $L= 372 \, \mathrm{\mu m}$).
 Full symbols: water (circle: $L=366 \, \mathrm{\mu m},v=20 \, 
\mathrm{\mu m/s}$, square:
$L=500 \, \mathrm{\mu m}, v=1 \,
 \mathrm{\mu m/s}$).
We observe that for $L> 300 \, \mathrm{\mu m}$ the  finite size effects are
negligible and the scatter of the data is mostly due to
the finite width of histogram bins.
In this way we have access to the large scale form of $\phi_{\text{exp}}(x)$.}
\label{f:gaussian2}
\end{figure}

In order to determine the experimental width distribution
we have employed the same experimental setup as  described in
Ref.~\cite{moulinet_contact_line};
the contact line  is imaged with a progressive
scan CCD camera equipped with a microscope.
 Two different
magnifications are used, corresponding to a pixel of size
  $\Delta x= 6.10$ and $\Delta x =2.14$ $\mathrm{\mu m}$.
After the analysis of the images we obtain a digitized line of $760$ pixels.
Thanks to the good contrast and signal-to-noise ratio of the CCD, the final
resolution is still one pixel.
The experimental lines are cut
into samples of size $L= n \Delta x$, where $n$ is
the number of pixels in the sample.
From each line we extract an ensemble of $(760-n)$ samples, whose
 width is simply given by  $w^2 = 1/L \sum_{i=1}^n u_i^2$.
These operations are repeated over all the configurations
 detected by the camera.
Clearly the values of $w^2$ are not independent\footnote{Let us
evaluate the number of independent configurations that can be
obtained in our experiment. Using, for example,  the lower
magnification  the observed length of the line  is
$4.7\,\mathrm{mm}$. During an experiment the total displacement is
of the order of $15\,\mathrm{mm}$.
 As the maximum width of the contact line is $\sim 30 \,\mathrm{\mu m}$,
the number
of independent configuration is 500.
 As an example, for a size  $L=500\,\mathrm{\mu m}$, $\phi$ is then determined
from typically $5000$ independent samples.}, but this procedure
ensures that no information is lost.

 The histogram derived by  the
data gives access to the universal function $\phi_{\text{exp}}(x)$ which is plotted in \fig{f:gaussian2} for various sizes $L$ and for various experimental
 conditions. For samples
of size in the range $ 300\,\mathrm{\mu m}<L< 800\,\mathrm{\mu m}
$, the obtained function $\phi_{\text{exp}}(x)$ is independent of the size, the
viscosity and the velocity of the receding meniscus. 
In smaller samples ($ 100\,\mathrm{\mu m}<L<
300\,\mathrm{\mu m} $), the distribution is sensitive to finite
size effect related to the details of the way the disorder is
created. Samples bigger than $L= 800\,\mathrm{\mu m}$ cannot be
treated due to the significant statistical noise.


At this stage, the universal function $\phi(x)$ which we have extracted from
experiments can
 be compared to the theoretical evaluation within the Gaussian
approximation.
 Previous theoretical works
 \cite{racz.random.94,racz.growth.94,plischke.curvature.94,rosso_width}
deal with samples where the  boundary conditions
were periodic, which is  obviously not
the case of our experiment (see \fig{f:cl}).
In order to calculate  $\phi(x)$ in the
Gaussian approximation for
 open  boundary conditions, we generalize the discussion of
Ref.~\cite{rosso_width}.
  The main difference, as it is briefly discussed in
Refs.~\cite{antal.1overf.01,ledoussal_Rcalcul}, lays  in the Fourier
decomposition of the
path $u(x)$.  In the case of open boundary conditions, the  general
path of size $L$ takes the form:
\begin{equation}
u(x)= \sum_{n=1}^{\infty} a_n \cos(\tilde q_nx),
\label{e:openFourier}
\end{equation}
where $\tilde q_n=\frac{\pi n}{L}$.
The probability associated with this path is:
\begin{equation}
\label{e:openGaussianGenN}
{\cal P}[u] = {\cal N} \prod_{n=1}^{\infty} e^{-1/2 a_n^2 G_n^{-1}},
\end{equation}
where ${\cal N}$ is the normalization factor and, for large $L$,
the exact disorder-averaged
2-point function $G_n$ takes the form  $ G_n \rightarrow  C/n^{1 + 2 \zeta}$.
The expression for $P(w^2)$ follows from the generating functions 
of the moments:
\begin{equation}
W(z) =  \int_{0}^{\infty} dw^2  P(w^2) e^{-z w^2}.
\label{e:generatrice}
\end{equation}

Similarly to   the case of periodic boundary conditions  \cite{rosso_width},
we write
\begin{eqnarray}
W(z)  &=& \prod_n
\left(\frac{1}{G_n^{-1} z +1}\right)^{1/2} \nonumber
\label{e:openWcalculation}
\end{eqnarray}
The function $\phi(x)$ is given by   $W(z)$ through
 an inverse-Laplace transform:
\begin{equation}
\phi(x) = \int_{- i \infty}^{+ i \infty} \frac{d z}{2 \pi i} e^{zx} \, \prod_{n=1}^{\infty}
\left( \frac{n^{\alpha} A}{z + n^{\alpha} A}   \right)^{1/2},
\label{e:openphi}
\end{equation}
with $A=\frac{1}{2}\sum_{n=1}^{\infty} \frac{1}{n^{\alpha}}$ and
$\alpha=1+2\zeta$.
This complex integral can be written as a sum of all the tadpole contributions:
\begin{equation}
\phi(x)= \sum_{n=0}^{\infty} \frac{ (-1)^n }{\pi}
  \int_{a_1(n)}^{a_2(n)}
dz e^{zx}  \prod_{m=1}^{\infty}
\left| \frac{m^{\alpha} A}{z + m^{\alpha} A}   \right|^{1/2},
\label{e:tadpole}
\end{equation}
with  $a_1=-A(2n+2)^{\alpha}$ and $a_2=-A(2n+1)^{\alpha}$.  The product
in \eq{e:tadpole} converges slowly, and several thousand terms need to   be
computed. Once this is done, however, only a few terms in the external sum
of \eq{e:tadpole} are sufficient to obtain $\phi(x)$ with high precision.

\begin{figure}[t]
\includegraphics[width=\columnwidth]{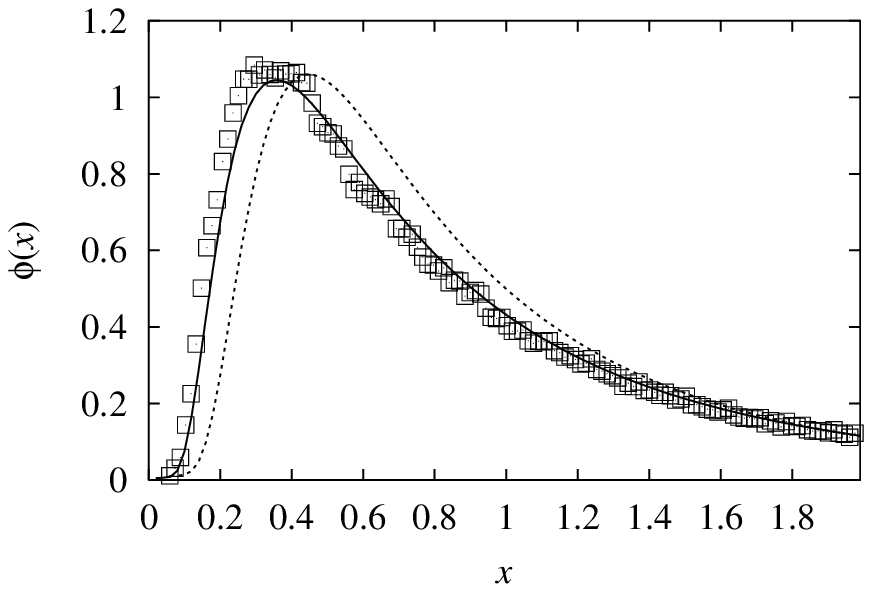}
\includegraphics[width=\columnwidth]{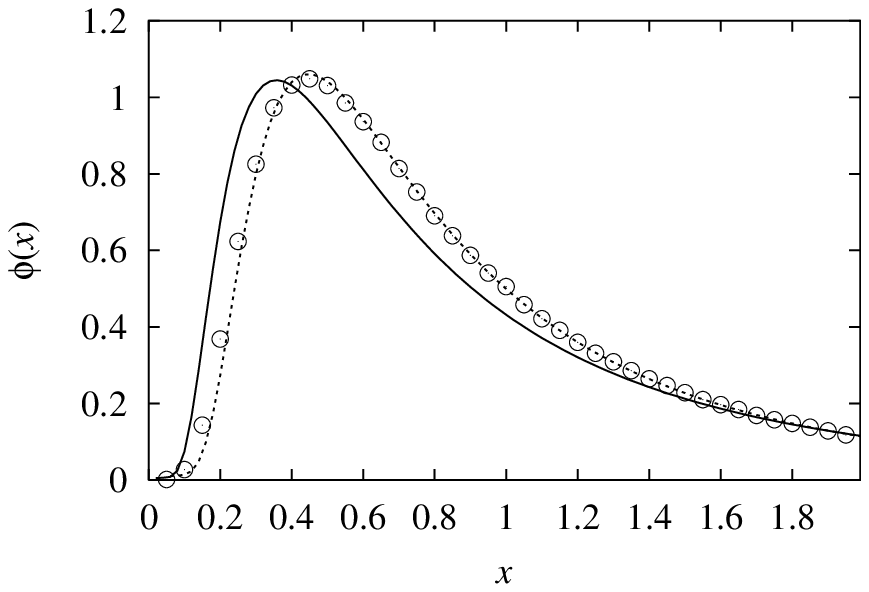}
\caption{Top: we compare the  scaling function $\phi_{\text{exp}}(x)$ obtained
from  the experimental data
(squares: water $L=500 \, \mathrm{\mu
m}, v=1 \,
 \mathrm{\mu m/s}$)   with
 the curves calculated by means of  \eq{e:tadpole}.
 We  notice the good agreement with the curve obtained for $\zeta =0.505$ (continuum line).
 Bottom: we compare  the  scaling function $\phi_{\text{num}}(x)$ obtained from the
 numerical  data   (circles: $L=32$, $10^6$ independent samples)  with the
 curves calculated by means of
 \eq{e:tadpole}.
 We  notice a perfect  agreement with the curve obtained for $\zeta =0.39$ (dashed line).}
\label{f:gaussian1}
\end{figure}

In Ref.~\cite{rosso_width} it is discussed how the width distribution of an interface
moving in a random medium, characterized by short range elastic interactions,
is described in an excellent way by the distribution
obtained within the Gaussian approximation.
For sake of completeness, we show the scaling function $\phi_{\text{num}}(x)$
computed from the numerical study  of \eq{motion}  at the
depinning threshold.
We use the algorithm from Ref.~\cite{rosso_krauth_longrange}
 and we calculate critical lines of
size $L=256$ with periodic boundary conditions.
$\phi_{\text{num}}(x)$ is obtained with the same procedure employed to deal with
experimental data, by using samples of size $L=32$. In this way, we
find the scaling function corresponding to the case of  open boundary conditions.

In \fig{f:gaussian1} we summarize our results.
As expected from  Ref.~\cite{rosso_width}, the function $\phi_{\text{num}}$
derived from the numerical study of  \eq{motion} is in perfect agreement 
with the function obtained in the Gaussian approximation with $\zeta = 0.39$.
The shape of $\phi_{\text{exp}}$ is also in good agreement with the calculated 
function, but it is clearly shifted with respect to $\phi_{\text{num}}$ and best approximated with $\zeta = 0.505$.


 To our knowledge, this kind of analysis is applied
for the first time to experimental data. It offers the possibility
to evaluate the roughness exponent from a unique value of $L$ and it
can be useful if the accessible range of scaling is small.
Moreover we have  detected finite size effects that are
invisible on the simple representation of $\overline{w^2}(L)$.
Due to the reduction of range of scaling, the analysis of $\overline{w^2}(L)$
yields $\zeta=0.52 \pm 0.04$, slightly larger than the previous determination  
 \cite{moulinet_contact_line}.


Our study confirms that a real contact line
cannot be described by  the Joanny-de~Gennes model.
The same type of equation of motion has been proposed to describe 
the propagation of crack fronts~\cite{gao_crack_elasticity}.
In this case   the experiments yield  $\zeta \sim 0.5 - 0.6$ 
\cite{schmittbuhl,deleplace}. Dynamical mechanism have been introduced 
recently \cite{bouchaud_bouchaud} to account for the anomalous value of 
$\zeta$ for crack fronts.
On the other hand, for the contact line, a precise analysis of
the motion justifies the quasi-static hypothesis\cite{moulinet_contact_line} 
 and no such dynamical mechanism has to be considered.
The study of the width distribution of crack fronts could help 
to understand if the two systems belong  to the same universality class and 
 to unravel the origin  of the discrepancy between the theoretical 
and the experimental values of $\zeta$.
In fact, in both cases, the derivation of the
 long-range elastic term 
in  \eq{motion} is obtained by
a development to  first order in the deformation.
This truncation   avoids the existence in the equation of motion 
 of  non-harmonic corrections which, as it has   been   shown for the 
 short-range elastic interactions
 \cite{rosso_krauth_non-harmonique}, can drastically change
the critical behavior of the interface. 
Moreover the presence  of  non-linear  terms
 has already  been introduced
in Ref.~\cite{golestanian} to understand the dynamics
of a contact line  in the high velocity limit. 
It would be very interesting to study the effect of these terms on the
 critical properties   at the depinning transition.  
Unfortunately, up to now  the numerical computation  of this  new  equation
 of motion  is impossible because of the presence of non-convex terms 
in the elastic energy.

We thank C.~Guthmann, P.~Le Doussal, J.~Vannimenus and J.~K.~Wiese 
for useful discussions.


\end{document}